# Detection Mechanism in SNSPD: Numerical Results of a Conceptually Simple, Yet Powerful Detection Model

Andreas Engel, Julia Lonsky, Xiaofu Zhang (张孝富), Andreas Schilling

*Abstract*—In a recent publication we have proposed a numerical model that describes the detection process of optical photons in superconducting nanowire single-photon detectors (SNSPD). Here, we review this model and present a significant improvement that allows us to calculate more accurate current distributions for the inhomogeneous quasi-particle densities occurring after photon absorption. With this new algorithm we explore the detector response in standard NbN SNSPD for photons absorbed off-center and for 2-photon processes. We also discuss the outstanding performance of SNSPD based on WSi. Our numerical results indicate a different detection mechanism in WSi than in NbN or similar materials.

## I. Introduction

SUPERCONDUCTING nanowire single-photon detectors (SNSPD) have seen tremendous improvements since their first implementation [1]. These improvements would not have been possible without an in-depth understanding of the whole detection process. This detection process may be divided into three steps: photon absorption, normal domain generation, and electronic detection of the absorption event. The first and the last steps of this process are well understood.

The absorption probability can be significantly improved by incorporating the superconducting meander into an optical cavity [2], [3] or by the use of plasmonic nanostructures [4], [5]. The overall system detection efficiency can be further improved by matching the meander geometry to the output of optical fibers and having a reliable self-alignment procedure of the fiber with respect to the detector [6], [7].

The discovery of an electro-thermal feedback between the read-out electronics and the detector [8] was an important step to understand the speed limitations of the detectors. New read-out schemes try to reach the ultimate limit in terms of maximum count-rates [9], [10] or to circumvent the problems that come with multipixel arrays [11],[12].

By contrast, the mechanism responsible for the formation of the initial normal domain following photon absorption has received limited attention until very recently. For a long time, the original proposal that the absorption of a photon creates a normal conducting hot-spot [1],[13], has been widely accepted, although it soon became clear that this model leads to inconsistencies when compared with experimental results [14].

Over the years several attempts have been made to describe the detection mechanism in SNSPD [14]-[17] with limited success, often due to gross simplifications necessary to obtain solutions. Very recently, we have proposed a simple detection model [18] by using established results wherever possible, and solving the resulting differential equations numerically with a common software package [19]. Here we will recapitulate the key features of our model, present some important improvements, and present a further verification using experimental results.

## II. Numerical Detection Model

In the proposed detection model [18] the detection process can be divided into three independent process steps, which we will discuss in the following. In general, we assume the superconducting films to be sufficiently thin to restrict the calculations to two dimensions. The origin of our coordinate system is in the centre of a rectangular strip, with the *x*-axis pointing along the length of the strip and in the direction of the current flow, and the *y*-axis in the transverse direction.

### A. Quasi-particle multiplication and diffusion

For the most important situations with regard to possible applications we have photon energies $h\nu \gg \Delta$, the superconducting energy gap per electron, but still less than the work function of the superconducting material. The process of quasi-particle (QP) multiplication has been studied in detail [20]. As already stated in the first paper about the detection mechanism in SNSPD [13], in most cases the time-dependence of the number of excess QP can be satisfactorily approximated by a simple exponential function. However, in [13] it was assumed that the concentration of QPs follows the solution of the diffusion equation in an extended two-dimensional film, simply multiplied by the total number of QPs. This approach neglects a possible spatial dependence of the creation of QPs.

Instead, we make the assumption [18] that the photon energy is transferred to one excited electron that starts diffusing in the superconducting strip with the diffusion coefficient $D_e$ of normal electrons. While the electron diffuses

This work was supported in part by the Swiss National Science Foundation under project number 200021_146887/1.
The authors are with the Physik Institut, University of Zurich, Switzerland (e-mail: andreas.engel@physik.uzh.ch).

away from the absorption site, it loses the excess energy and creates excess QPs with an overall efficiency $\zeta < 1$. The local rate of QP creation is assumed to be proportional to the probability density $C_e(\vec{r}, t)$ of the first excited electron. This process of QP-multiplication happens on a very short time-scale $\tau_{qp}$. After their creation, the QPs themselves are subject to diffusion with a diffusion coefficient $D_{qp} < D_e$, and the QP density $C_{qp}(\vec{r}, t)$ also follows the diffusion equation in two dimensions. Eventually, these excess QPs recombine to form Cooper-pairs on a time-scale $\tau_r \gg \tau_{qp}$. The whole process of QP creation and diffusion is described by a set of coupled differential equations [18],

$$\frac{\partial C_e(\vec{r},t)}{\partial t} = D_e \nabla^2 C_e(\vec{r}, t), \tag{1}$$

$$\frac{\partial C_{qp}(\vec{r},t)}{\partial t} = D_{qp} \nabla^2 C_{qp}(\vec{r}, t) - \frac{C_{qp}(\vec{r},t)}{\tau_r} + \frac{\zeta h \nu}{\Delta \tau_{qp}} e^{-\frac{t}{\tau_{qp}}} C_e(\vec{r}, t). \tag{2}$$

*B. Current redistribution*

Suitable superconducting materials for SNSPD are typically strongly type-II with a large magnetic penetration depth $\lambda_{GL}$. With the thickness of the films $d \ll \lambda_{GL}$, this results in an even larger effective penetration depth $\Lambda = 2\lambda^2_{GL}/d \gg w$, the width of the meander strips. This ensures a highly uniform current density along the straight sections of the meander in the equilibrium situation.

After photon absorption, the current density will no longer be uniform in the vicinity of the absorption site due to the inhomogeneous distribution of QPs. For NbN and similar materials we estimated the Ginzburg-Landau (GL) relaxation time $\tau_{GL} \lesssim 1$ ps [21]. The QP multiplication and diffusion happens on a time scale of a few picoseconds. Therefore, we make two assumptions allowing us to use stationary equations: *i)* the QPs are always in a local near-equilibrium such that the local increase in the QP density equals the reduction in the superconducting electron density $n_{se}$, and *ii)*, the redistribution of the bias current is instantaneous and can be calculated from the respective density of superconducting electrons.

In our original model, we set the current density in the *x*-direction $j_x(\vec{r}, t) \propto n_{se}(\vec{r}, t)$ and required current continuity $\vec{\nabla} \cdot \vec{j} = 0$. This approximation kept our model linear, but would not reproduce the expected current-crowding effect near the equator of a hole in a superconducting strip, for example [16], [22]. A more realistic current distribution can be calculated from the relation between the drift velocity of superconducting electrons $v_s$ and the phase $\varphi$ of the superconducting condensate $\vec{v}_s = \frac{\hbar}{m} \vec{\nabla} \varphi$, with $\hbar$ the reduced Planck constant and $m$ the electron mass. Still requiring current continuity, we have to solve [23]

$$\vec{\nabla} \cdot (n_{se} \vec{\nabla} \varphi) = 0. \tag{3}$$

Additionally, we take now into account that the relation between drift velocity and current density is no longer linear for current densities approaching the critical drift velocity $|\vec{v}_{s,c}|$ corresponding to the critical current density $|\vec{j}_c|$. High drift velocities have a pair-breaking effect that leads to a reduction of $n_{se}$ [24]:

$$n_{se}/n_{se,0} = 1 - \frac{\left(|\vec{v}_s|/|\vec{v}_{s,c}|\right)^2}{3}, \tag{4}$$

with $n_{se,0}$ the density of superconducting electrons for zero applied current. As a consequence, the problem of calculating the current distribution becomes non-linear. The original, linear results turn out to be still a good approximation to the more precise solutions obtained with (3) and (4), for the situations considered in [18], i.e. the absorption of a single photon in the center of the strip. This is because we can use the smooth and continuously varying densities of superconducting electrons derived from solving (1) and (2), which result in only a weak current-crowding effect. However, for some of the situations we will discuss below, e.g. absorption of a photon close to the edge of the strip, it is important to solve the non-linear equations (3) and (4) to obtain realistic results.

*C. Normal-domain trigger mechanism*

In [18] we have shown that photons of a given energy lead to the entry of vortices from the edges of the strip and the subsequent formation of a normal-domain at lower bias currents and well before the two other considered mechanisms, the hot-spot model [13] and the QP model [14], can trigger a normal-domain. And because the present improved calculation of the current redistribution does not change the trigger condition for them, we will not consider these other detection mechanisms here.

The vortex trigger-mechanism is based on the suppression of the edge barrier for vortex-entry due to an increase of the current density and a reduction of the superconducting electron-density. Such a photon-assisted vortex-entry was first suggested in [15]. In the previous model, we calculated the effective vortex potential by explicitly determining the force on a test vortex as function of position, followed by the calculation of the line integral from the point of entry to a position inside the strip.

Now, we use a computationally less demanding approach as suggested in [22]. This approach also uses the London-model of a vortex with a normal-conducting core of radius $\xi$, the superconducting coherence length, surrounded by the circulating screening current. Within this model the vortex self-energy may be written as $E_{\text{self}}(\vec{r}) = \phi_0 I_{\text{circ}}(\vec{r})/2$, with $I_{\text{circ}}(\vec{r})$ the total self-generated current circulating around the core of a vortex sitting at position $\vec{r}$, and $\phi_0 = h/2e$ the magnetic flux quantum. We calculate $I_{\text{circ}}(\vec{r})$ by putting the vortex core at position $\vec{r} = (x_0, y_0)$ and considering only the half-plane with $x \geq x_0$. The phase change for one loop around the core has to be $2\pi$. We set the phase $\varphi = 0$ for $-w/2 \leq y < y_0 - \xi$ and $\varphi = \pi$ for $y_0 + \xi < y \leq w/2$ and apply (3) to obtain the current density $j_{\text{circ}}$ for the circulating current, $I_{\text{circ}}$ is obtained by integration of $j_{\text{circ}}$.

TABLE I
MATERIAL PARAMETERS USED IN THE SIMULATIONS

| Material | Energy gap (meV) | Coherence length (nm) | Penetration depth (nm) | Normal-state diffusion coefficient (nm$^2$/ps) | Time constant $\tau_{qp}$ (ps) | Time constant $\tau_r$ (ps) |
|---|---|---|---|---|---|---|
| NbN | 2.3 | 4.2 | 430 | 52 | 1.6 | 1000 |
| TaN | 1.3 | 5.2 | 520 | 60 | 1.6 | 1000 |
| WSi | 0.53 | 8.0 | 1400 | 75 | 1.6 | 1000 |

Zero temperature material parameters used for the simulations. Except for the time constants the parameters have been experimentally determined from magneto-conductivity measurements. Due to the lack of experimental data, the time constants are assumed to be the same for all three materials.

The work done by the source of the applied current to move the vortex from its point of entry to the position $\vec{r}$ can be calculated as $W_I(\vec{r}) = \phi_0 \Delta I(\vec{r})$, with $\Delta I(\vec{r})$ being that part of the applied current that flows between the point of entry and position $\vec{r}$ of the vortex. The Gibbs free energy, which is equivalent to the potential energy of the vortex, is then simply given by $U(\vec{r}) = E_{\text{self}}(\vec{r}) - W_I(\vec{r})$. We have verified that this new algorithm reproduces the analytical solutions for the case of homogeneous current densities, as expected.

The simple London-model of a vortex loses its accuracy for vortex distances to the edge comparable to $\xi$, and becomes invalid for distances less than $\xi$. However, it has been successfully applied in many different situations [25]-[27]. These limitations of the London model may result in a quantitative discrepancy between calculated and experimental vortex-entry currents. Numerical results may need adjustment of the vortex entry current, e.g. for the case when no photon is absorbed. Another possibility to fit numerical results to measured data is the efficiency $\zeta$, which serves as an adjustable parameter in our model.

There is at least one more possible mechanism that could lead to the formation of a normal-domain, which is based on the formation and subsequent unbinding of a vortex-antivortex pair [16]. In many respects this process is comparable to the vortex-entry model considered here and may be a competing mechanism for certain situations, e.g. in cases when indeed a true normal-conducting core is formed that leads to a pronounced current-crowding effect. The newest results [28], [29] within this vortex-antivortex model appear to be very similar to the results obtained with our vortex model, however.

*D. Input parameters*

The current implementation of our simulation algorithm requires the input of certain material, device specific, and other experimental parameters. Particularly important are realistic material parameters; a complete list of required parameters is given in Table I for the three materials used in this publication: NbN [30], TaN [31], and WSi [32]. Except for the time constants, the zero-temperature values have been derived from magneto-conductivity measurements. We assume the superconducting parameters to be temperature dependent and the diffusion coefficient for QPs is derived from the normal-state diffusion coefficient, for details see appendix in [18].

The thermalization time for NbN has been experimentally determined to $\tau_{th} \approx 7$ ps near $T = 4$ K [33]. The time constant $\tau_{qp}$ has been chosen to result in a slightly longer thermalization time of about 10.5 ps. The recombination time constant $\tau_r$ has been chosen as an average value [34]. However, the results depend only very weakly on the exact value of $\tau_r$. For the other two materials, TaN and WSi, we are not aware of experimental data on these two time constants. For this reason we tentatively use the same values as for NbN. The conversion efficiency $\zeta$ is most likely also material dependent. In our model it is used as an adjustable parameter and is set to $\zeta = 0.25$ for the simulations.

The device parameters are the strip width $w$ and the thickness $d$ of the superconducting strip. The length of the strip is set to $L = 1$ μm for all simulations. Furthermore, we can select the following experimental parameters: the photon wavelength, the photon absorption position, the temperature, and the bias current. The photon wavelength and the bias current are usually varied over typical experimental ranges.

### III. SIMULATION RESULTS

Before we present our new results we demonstrate that the changes we have made in the algorithm still lead to the same results as presented in [18]. The key findings were the linear relation between threshold current and photon energy, and that a linear extrapolation to zero photon energy coincides with the vortex-entry current, i.e. the current for which the edge barrier for vortex-entry vanishes when no photon is absorbed.

In Fig. 1 we show simulation results (red open data points) obtained with the new algorithm but otherwise identical parameters together with experimental data from [31] (black filled data points) for a TaN SNSPD. The simulation data shown are limited to the low energy part of the considered photon spectrum $h\nu \lesssim 1$ eV. It turned out that the algorithm used to calculate the current distributions from the nonlinear differential equations (3) and (4) cannot adequately handle situations for which the photon absorption leads to a strong suppression of the density of superconducting electrons. However, for the range of photon energies for which the calculations lead to consistent and physically meaningful results, the threshold currents that trigger a detection event are to a very good approximation linear in photon energy.

The red solid line in Fig. 1 is a linear extrapolation of the simulation data to higher photon energies. It already gives a fair description of the experimental data. The red dashed line is a linear fit to the experimental data. We can even obtain the same linear behaviour by further adjusting the vortex-entry current that is obtained from the intersection point of the linear extrapolation with the ordinate to that value derived from the linear extrapolation of the experimental data. This would require a change from $0.79\,I_c$, where $I_c$ is the depairing critical-current, to $(0.83 \pm 0.02)\,I_c$, which seems to be very reasonable taking into account the expected loss of accuracy of the London-model for vortices very close to the edge. The slope of our simulation data can be easily adjusted to the experimental best fit by a change in $\zeta$, as detailed in [18]. The best fit is obtained for $\zeta = 0.32 \pm 0.04$, which still lies well within the expected range [14], [31].

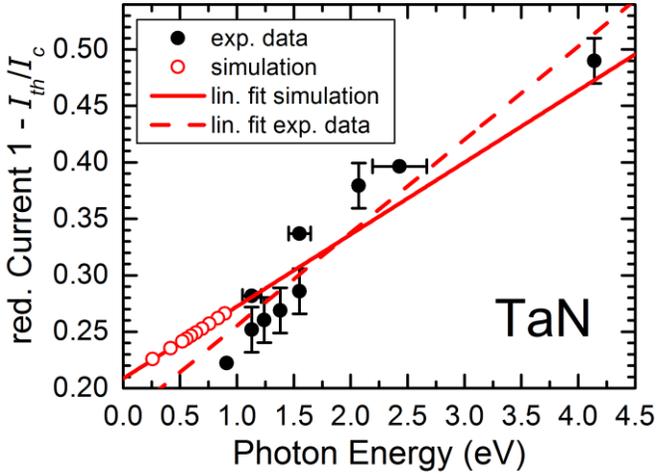

Fig. 1. The reduced threshold current $1 - I_{th}/I_c$ as function of the photon energy from simulations is compared with experimental results. The same TaN device parameters were used as in [18]. The solid red line is a linear fit to the simulation data, the dashed line is a fit to the experimental data.

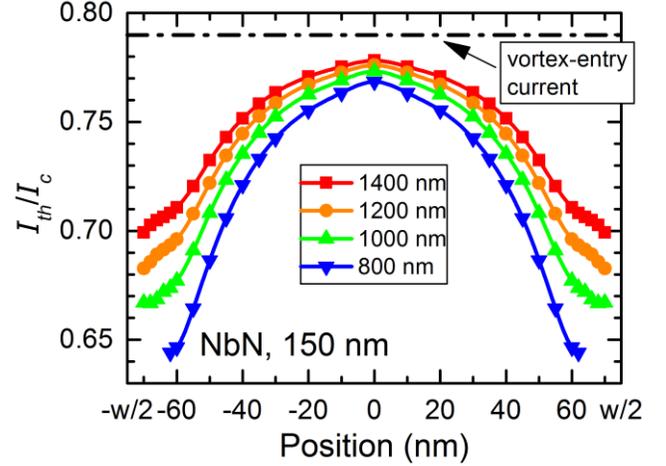

Fig. 2. Threshold current $I_{th}/I_c$ plotted vs. the absorption position for different wavelengths as indicated using NbN material parameters and a strip width $w = 150$ nm. The vortex-entry current which equals the experimental critical current in the vortex model is also indicated.

In [18] we also had a look at the times, when the normal-domain is triggered. There we have found that the vortex model leads to the formation of the normal-domain within less than $\approx 2.5$ ps or $\approx 0.25\,\tau_{th}$. The results obtained with the present nonlinear model suggest that the normal domain at the threshold current is triggered at $\approx 0.5\,\tau_{th}$. Although our nonlinear algorithm results in certain small adjustments of the numerical results as compared to the previous linear model, the qualitative behaviour is the same, including the clear result that vortices are much more likely to trigger a normal domain than the alternative mechanisms suggested in [13] and [14].

*A. Position-dependence*

There have been experimental indications [35] that the detection probability may depend on the position of absorption of the photon. However, most detection models so far considered absorption in the centre of the superconducting strip, only. We have systematically studied the dependence of the threshold current on the photon absorption position. In Fig. 2 we show representative results obtained for a $w = 150$ nm wide, $d = 4$ nm thick NbN device at an operating temperature $T/T_c = 0.1$. The adsorption position was varied across the strip with a minimum distance to the edge of about one coherence length.

We find $\gtrsim 10\%$ larger threshold currents for absorption events near the centre of the strip than near the edges. This trend is more pronounced for photons of higher energy. At a given absorption position, however, the threshold current is always a linear function of the photon energy. This effect can, at least partly, explain the typically observed rounding of the detection efficiency near, but above the threshold current.

Studying the simulation results in more detail also reveals the origin of this effect. Starting in the centre of the strip and moving the absorption site to one edge, at first an increase in the current density near the edge is responsible for the reduction of the threshold current. As the absorption site comes to within about 30 nm of the edge, the current density becomes smaller, which is overcompensated by a strong reduction of the superconducting order parameter near the edge leading to a much smaller vortex self-energy.

*B. Two-photon absorption*

Very strong experimental verification of the linear relation between threshold current and photon energy has been coming from a series of experiments applying the technique of detector tomography [36], [37]. A very interesting result of these experiments has been the observation that the total deposited energy is relevant, irrespective whether this energy is coming from a single photon or $n$ photons, absorbed spatially and temporarily within a certain limit, each with $1/n$ of the energy of the single photon.

For two or more photons absorbed at precisely the same point in space and time, this observation is not surprising, if the linear equations describing the QP multiplication and diffusion are correct. In pulsed laser experiments with pulse widths of just a few picoseconds or less, the temporal difference for a 2-photon event is usually small enough to be considered simultaneously, because one single photon reduces the edge-barrier for vortex-entry within a few picoseconds and the edge-barrier will remain low, close to the minimum value for a time span of at least about the same duration. However, spatial distances between two absorbed photons can be large enough to be considered independent 1-photon events.

We have made a first attempt to understand 2-photon events by varying the distance between two simultaneously absorbed photons at $x = 0$ and varying $y$-positions, and at $y = 0$ and increasing the distance along the $x$-coordinate, respectively. Simulations were done for the same NbN device as in Section

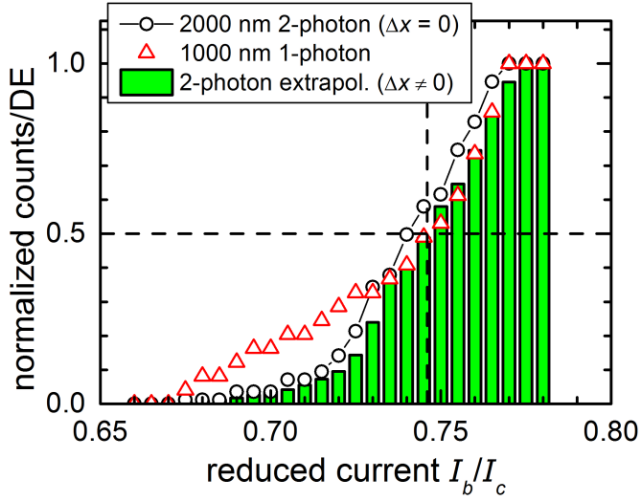

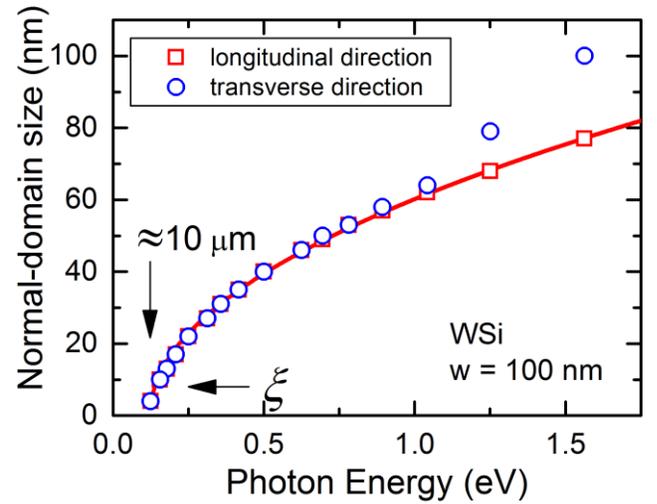

Fig. 3. Comparison of three different histograms as indicated. The 1-photon 1000 nm data are derived from the data in Fig. 2. The 2000 nm 2-photon data for $\Delta x = 0$ are extracted from simulations with varying $y$-positions of the two photons. The extrapolated data have been derived from a limited dataset with $\Delta x \neq 0$. The dashed lines mark a 50%-detection criterion and the corresponding bias current.

Fig. 4. Size of the normal-core in $x$- and $y$-direction for photons of different energy absorbed in $w = 100$ nm wide WSi strip. The red line is a least-square fit to a square-root relation between the photon energy and the diameter of a normal-conducting spot. Transverse extension of the normal-core starts to deviate at $\approx 1$ eV from this simple relation due to edge effects.

III.A and photon wavelengths $\lambda = 2000$ nm to be compared with 1-photon events with $\lambda = 1000$ nm.

Already the simple case of $x = 0$ and independent $y$-positions for two photons requires a 2-D mapping of absorption positions. This has been done for a step-size $\Delta y = 10$ nm. Fig. 3 shows the resulting cumulative histogram of $I_{th}/I_c$, which should be a representation of normalized detection efficiencies (DE) versus reduced bias currents typically measured in experiments. In [36], [37] threshold currents were determined for a fixed DE, e.g. DE = 0.5. If 1-photon and $n$-photon events are indeed equivalent, a given DE should be reached for the same reduced bias currents. Comparing the histograms we see immediately that this is generally not the case, e.g. for the bias current that results in 50% DE for 1000 nm 1-photon events, the 2000 nm 2-photon events would be detect with almost 60% probability.

However, we now consider the case with events for which $\Delta x \neq 0$. A complete sampling of 2-photon events is computationally very time consuming. Instead, we calculated threshold currents for events with $y = 0$ for both photons and $\Delta x \neq 0$, and used these results to scale the 2D-mapping that we have obtained for the case $\Delta x = 0$. This already gives a very good agreement between the normalized counts or DE for 1-photon and 2-photon events $\gtrsim 0.4$. The discrepancy at lower DE is not surprising, because the extrapolation that we apply does not adequately represent 2-photon events for which both photons are absorbed close to the edge, and those events result in particularly low threshold currents. Overall, the results so far are strong evidence that the vortex-model can reproduce the experimental 2-photon results from detector tomography.

### C. WSi

In the last few years a number of alternative materials to NbN have been suggested to be used in SNSPDs, such as TaN [31] or NbTiN [38], which may offer some advantages for certain situations, although their overall performance was very similar to NbN based SNSPD. The situation is somewhat different for a recently developed device based on amorphous WSi [39], [7]. The amorphous structure of the superconducting film is advantageous from a technological point-of-view, and the comparably low $T_c$ significantly increases the detection efficiencies for low-energy photons. These advantages come at the cost of lower operating temperatures and lower signal-to-noise ratio of the detector signal.

Compared to SNSPD based on NbN and related materials, there are also qualitative differences in the WSi detector characteristics. For example, the plateau of constant DE for $I_b > I_{th}$ can be significantly larger (depending on operation temperature), it is very flat, and $I_{th}$ is better defined due to reduced rounding in the vicinity of $I_{th}$ [7]. These features are difficult to understand based on the amorphous structure and the lower $T_c$ alone.

We prepared a series of $W_xSi_{1-x}$-films ($0.2 \leq x \leq 0.3$) and characterized their superconducting and normal-state properties from magneto-conductivity measurements (32). In Table I we list the important material parameters with approximate values as we expect them for SNSPD. While $T_c$ (and along with it the energy gap) is reduced by a factor ~4 as compared to NbN, the magnetic penetration depth is ~3 times larger. The larger penetration depth is a consequence of a much lower density of states at the Fermi-energy of WSi, which results in a much lower density of superconducting electrons as compared to NbN. As a consequence, the condensation energy density is significantly lower in WSi than in NbN.

We performed simulations for WSi parameters and we assumed a width $w = 100$ nm, a thickness $d = 4$ nm, and an operating temperature $T/T_c = 0.2$. Already the results of the QP multiplication and diffusion simulations clearly indicate that even infrared photons produce large normal-conducting areas, sometimes large enough to span the complete width of the superconducting strip, resulting in a complete cross-section becoming normal-conducting even without applied current.

In Fig. 4 we plot the extension of the normal-conducting core in the longitudinal and transverse directions, respectively. The solid line is a square-root fit to the data in the longitudinal direction. From these data we can determine a minimum photon wavelength of ≈10 μm that leads to a normal-conducting core of size ξ, to be compared with a value well below 1 μm for a comparable NbN detector. At a corresponding wavelength of 800 nm, the transverse extension of the normal-core is 100 nm and spans the entire strip.

The current implementation of our algorithm does not allow for a detailed study of the threshold current in such situations, but the QP multiplication and diffusion results suggest that it is this relatively large normal-core that triggers the detection event. Our studies clearly suggest a different triggering mechanism in WSi based SNSPD as compared to detectors based on NbN and related materials.

## IV. Conclusion

We have presented recent results obtained with an improved algorithm of our photon detection model for SNSPDs. This algorithm allows for the calculation of realistic current distributions, provided that the reduction in the density of superconducting electrons is not too strong. These new results confirm the strictly linear relation between photon energy and detection threshold current, also in situations when the photon is absorbed close to one edge. Photons absorbed close to the edge are actually detected with a lower threshold current, which has been suggested previously.

Within our vortex detection model we have confirmed results obtained with detector tomography, namely that only the total photon energy is important, independent of how many photons are involved. We are confident to be able to confirm it also for lower DE doing a complete simulation at least for 2-photon events.

Using experimentally determined parameters for amorphous WSi-films we could show that near-infrared photons can already lead to large normal-conducting areas, even without the application of a bias current. This indicates a different mechanism for the formation of the initial normal-domain than in NbN based devices, and could explain some of the remarkable properties of WSi SNSPD.

All of the above results demonstrate the power of our detection model and it may be useful to explain and understand a range of other details connected to the detection mechanism in SNSPD.

## Acknowledgment

The authors would like to thank J. Renema for fruitful discussion, M. Frick for valuable support with programming and D. Y. Vodolazov for pointing out a way of calculating more accurate current distributions.